\documentclass{article}
\usepackage{spconf,amsmath,epsfig}

\usepackage{bm}
\usepackage{algorithm}
\usepackage{multirow}
\usepackage{amsmath,amsfonts,amssymb,graphicx}

\def\x{{\mathbf x}}
\def\L{{\cal L}}
\makeatletter
\def\hlinewd#1{%
  \noalign{\ifnum0=`}\fi\hrule \@height #1 \futurelet
   \reserved@a\@xhline}
\makeatother
\begin{document}\sloppy

% Example definitions.
% --------------------
\def\x{{\mathbf x}}
\def\L{{\cal L}}

% Title.
% ------
\title{Deep Vocoder: Low Bit Rate Compression of Speech with Deep Autoencoder}
%
% Single address.
% ---------------
\name{Gang Min{\small $~^{1}$}, Changqing Zhang{\small $~^{1}$}, Xiongwei Zhang{\small $~^{2}$}, Wei Tan{\small $~^{1}$}\thanks{This work is partially supported by Natural Science Foundation of China(61701535) and Natural Science Foundation of Shanxi Province (2017JQ6033).}}
\address{$^{1}$\,Institute of Information and Communication, National University of Defense Technology, Xi'an, China\\
$^{2}$\,Army Engineering University of PLA, Nanjing, China\\
\{mgxaty, xwzhang\}@gmail.com, \{zhangcq1108, dzxxlab\}@163.com}

\maketitle

\begin{abstract}
Inspired by the success of deep neural networks (DNNs) in speech processing, this paper presents Deep Vocoder, a direct end-to-end low bit rate speech compression method with deep autoencoder (DAE). In Deep Vocoder, DAE is used for extracting the latent representing features (LRFs) of speech, which are then efficiently quantized by an analysis-by-synthesis vector quantization (AbS VQ) method. AbS VQ aims to minimize the perceptual spectral reconstruction distortion rather than the distortion of LRFs vector itself. Also, a suboptimal codebook searching technique is proposed to further reduce the computational complexity. Experimental results demonstrate that Deep Vocoder yields substantial improvements in terms of frequency-weighted segmental SNR, STOI and PESQ score when compared to the output of the conventional SQ- or VQ-based codec. The yielded PESQ score over the TIMIT corpus is 3.34 and 3.08 for speech coding at 2400 bit/s and 1200 bit/s, respectively.
\end{abstract}
\begin{keywords}
Deep Vocoder, speech coding, vector quantization, analysis-by-synthesis
\end{keywords}
\section{Introduction}
\label{sec:intro}

There is a great deal of interest in low bit rate compression of speech for its widespread use in both secure and satellite communications, however, it remains an open challenge, especially in the presence of background acoustic noises. In the traditional source-filter speech coding framework, speech encoding parameters including linear prediction coefficients and pitch are sensitive to environmental noises, which leads to degradation of speech quality inevitably in noisy conditions. Thus, many efforts have been made towards alternatives to the popular linear prediction coding model, such as the phase vocoder \cite{BSTJ:Flanagan}, \cite{TSA:Laroche}, multiband codec \cite{JSAC:Cox}, \cite{TASS:Griffin}, MFCC codec \cite{TASLPLow:Boucheron}, \cite{SPL:Gang}, artificial neural networks based codec \cite{ICASSP:Bengio}, \cite{ICC:Morishima}.

The last decade has witnessed great success of deep neural networks (DNNs), which helped to improve performance dramatically in various applications, such as automatic speech recognition, text-to-speech, supervised speech separation, \emph{et al}. Yet, DNNs are relatively less exploited in the field of lossy speech compression. Though a small amount of literatures proposed to use artificial neural networks for speech coding, the performance of which is difficult to compare with state-of-the-art vocoders due to the weak capability of early shallow neural networks for speech analysis and synthesis \cite{ICASSP:Bengio}, \cite{ICC:Morishima}.

Recently, deep autoencoder (DAE) with a binary coding layer was proposed for coding speech spectrograms \cite{INTERSPEECH:Deng}, \cite{IJEC:Jiang}, which opens up a new promising direction for compressing speech signal with DNNs. However, there are still limitations to be overcome, such as bit allocation and speech reconstruction. Essentially, the spectrogram coding methods mentioned above are relatively simple scalar quantization (SQ) methods for the latent representing features (LRFs) learned with DAE, whose performance is limited because SQ cannot remove the redundancy among LRFs vector components \cite{TIT:Gray}.

Different from the phonological recognition and synthesis- or wavenet-based low bit rate speech coding method \cite{TASLP:Cernak}, \cite{ICASSP:Kleijn}, this letter presents Deep Vocoder, a direct end-to-end speech compression method which uses DAE for speech analysis and synthesis. In \cite{INTERSPEECH:Deng}, LRFs in the coding layer of DAE are directly quantized to be either zero or one using SQ technique, here we propose to use the analysis-by-synthesis vector quantization (AbS VQ) technique with perceptual distortion criterion to encode LRFs efficiently, which is shown to provide a much better speech quality. Motivated by the human's auditory properties \cite{TASSP:Ephraim} and analysis-by-synthesis (AbS) technique which is broadly used in low bit rate codec \cite{ASSP:Kroon}, \cite{JSAC:Kroon}, AbS VQ changes the objective of vector quantization (VQ) of LRFs as: the codeword of minimum log-spectral reconstruction distortion is selected as the quantized LRFs vector. The conventional SQ or VQ approach quantizes LRFs vector itself directly in an open-loop fashion, however, the AbS VQ approach strategically uses a closed-loop technique known as analysis-by-synthesis. The synthesis stage employs DAE to reconstruct speech spectra for measuring the effect of quantization of LRFs on the final speech quality, and the analysis stage is performed followed by the synthesis step to select an appropriate codeword to minimize the log-spectral distortion between the original and reproduced speech signal. After the quantization procedure of LRFs vector, the speech waveforms is finally reconstructed from spectrogram by the well-known Griffin-Lim algorithm \cite{TASSP:Griffin}. To the best of our knowledge, DAE has not been used with AbS VQ for speech compression before, which yielded perceptual evaluation of speech quality (PESQ) \cite{ICASSP:Rix} scores competitive with state-of-the-art vocoders, such as the enhanced mixed-excitation linear predictive (MELPe) codec \cite{TSAP:McCree}, \cite{NATO:Melpe}.
\begin{figure*}[htp]
\centering
\includegraphics[width=5.2in]{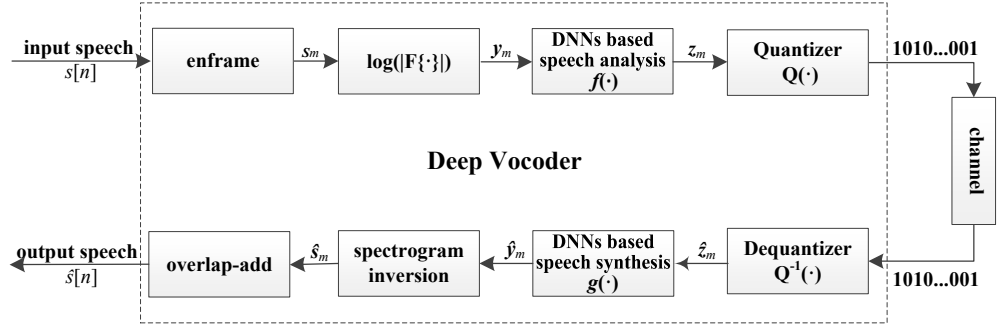}
% where an .eps filename suffix will be assumed under latex,
% and a .pdf suffix will be assumed for pdflatex; or what has been declared
% via \DeclareGraphicsExtensions.
\caption{Overview of Deep Vocoder.}
\label{fig_deepvocoder}
\end{figure*}
\section{Algorithm}
\subsection{ Overview of Deep Vocoder}
As is shown in Fig. 1, Deep Vocoder firstly enframes the speech waveforms \emph{s}[\emph{n}] by a window $w\left[n\right]$,
\begin{equation}
{s_m}\left[n\right]=s\left[{mR+n}\right]w\left[n\right]
\end{equation}
where $L (0\leq{n}\leq{L-1})$ is the window length, \emph{R} is the frame shift, $m(m=1,2,...,M)$ is the frame index. Then, The speech frame can be concisely denoted as,
\begin{equation}
{\bm{s}_m}={\left[ {{s_m}\left( 0 \right),{s_m}\left( 1 \right),...,{s_m}\left( {L - 1} \right)} \right]^{\intercal}}
\end{equation}

The log-magnitude spectrum of each speech frame is,
\begin{equation}
{\bm{y}_m}=\log(\left|{{\rm{F}}\left\{{\bm{s}_m} \right\}}\right|)
\end{equation}
where ${\rm{F}}\left\{{\bm{s}_m}\right\}$ is the \emph{N}-point fast Fourier transform (FFT) of $\bm{s}_m$, $\left|\cdot\right|$ denotes the modulus of a complex number. Due to the symmetry, the latter $N/2-1$ elements of $\bm{y}_m$ will be discarded.

In Deep Vocoder, we will introduce a speech analysis function \emph{f}, a speech synthesis function \emph{g}, and a quantizer \emph{Q},
\begin{equation}
f:{\mathbb{R}^{N/2 + 1}}\to{\mathbb{R}^K},g:{\mathbb{R}^K}\to{\mathbb{R}^{N/2 + 1}},Q:{\mathbb{R}^K}\to\left[{0,1}\right].
\end{equation}

At the encoder, the LRFs vector $\bm{z}_m\in\mathbb{R}^K$ is then learned through $f$ and it should be quantized as bit stream $b_m$,
\begin{equation}
{{\bm{z}}_m} = f\left( {{{\bm{y}}_m}} \right),{b_m} = Q\left( {{{\bm{z}}_m}} \right).
\end{equation}

At the corresponding decoder, $b_m$ is decoded and the log-magnitude spectrum ${\bm{\hat y}}_m$ is reconstructed through $g$,
\begin{equation}
{{\bm{\hat z}}_m} = {Q^{ - 1}}\left( {{b_m}} \right),{{\bm{\hat y}}_m} = g({{\bm{\hat z}}_m}).
\end{equation}

Generally, Deep Vocoder aims to optimize the tradeoff between using a small number of bits to compress speech signal and having small speech distortion,
\begin{equation}
\mathop {{\rm{min}}}\limits_{f,g,Q} \underbrace {\sum\limits_{m = 1}^M {d\left( {{{\bm{y}}_m},{{{\bm{\hat y}}}_m}} \right)} }_{{\rm{Distortion}}} + \lambda \underbrace {\sum\limits_{m = 1}^M {{{\log }_2}\left( {{b_m}} \right)} }_{{\rm{Number~of~bits}}}
\end{equation}
here, $\lambda$ controls the tradeoff and $d$ measures the distortion introduced by speech compression and decompression. This distortion measure is perceptually meaningful since it is defined in the log-spectral domain. The next sections will describe how to design $f$, $g$, and $Q$ in detail.
%\IEEEpubidadjcol
\subsection{Speech analysis and synthesis with DAE}
Establishing a model for speech analysis and synthesis is the basis of low bit rate speech compression, its main goal is to extract feature parameters for speech compression. DAE is a special type of DNNs, where the output layer has the same number of nodes as the input layer, and with the purpose of reconstructing its own inputs as similar as possible. As an important unsupervised learning model, DAE aims to build a hopefully simpler representation for a set of data, so it could be employed for accurately modeling speech spectrum and discovering high-level features for speech processing. DAE usually consists of an encoder and a decoder, where the encoder maps the input to a latent space and the decoder maps it back to the input space. As is shown in Fig. 2, it usually involves two phases to train a DAE model for speech spectrum, i.e., unsupervised pre-training and supervised fine-tuning. After the DAE model is trained, we can use the the encoder of DAE to analyze speech spectrum and extract the LRFs vector on the one hand, we can also use the decoder of DAE to synthesize speech spectrum from the LRFs vector on the other hand.

\begin{figure}[htp]
\centering
\includegraphics[width=3.1in]{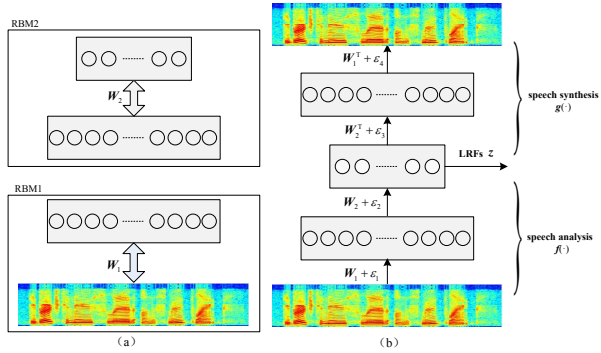}
% where an .eps filename suffix will be assumed under latex,
% and a .pdf suffix will be assumed for pdflatex; or what has been declared
% via \DeclareGraphicsExtensions.
\caption{Speech analysis and synthesis with DAE. (a) Illustration of pre-training for DAE; (b) Illustration of fine-tuning for DAE.}
\label{fig_speechAnalysisSynthsis}
\end{figure}
\subsection{Analysis-by-synthesis vector quantization for LRFs vector}
Quantization of speech coding parameters is a crucial step for compressing speech signal. In Deep Vocoder, speech coding parameters are solely LRFs vector extracted by DAE. Different from the conventional SQ or VQ method, here we propose to use the AbS VQ technique to quantize LRFs vector efficiently. As is shown in Fig. 3, the AbS VQ technique consists of two steps: a synthesis step that reconstructs speech spectrum from the codeword $\tilde{\bm{z}}_m$ and an analysis step that calculates the log-spectral distortion between the original speech spectrum $\bm{y}_m$ and the reconstructed speech spectrum $\hat{\bm{y}}_m$. These two steps will be repeated until the whole AbS VQ codebook $\mathcal{Z}$ is searched. Finally, the codeword with minimum $d(\bm{y}_m,\hat{\bm{y}}_m)$ is selected as the quantized LRFs vector. It is worth mentioning that the AbS VQ codebook $\mathcal{Z}$ is usually trained on the large-scale corpus using the LBG algorithm.
\begin{figure}[h]
\centering
\includegraphics[width=3.1in]{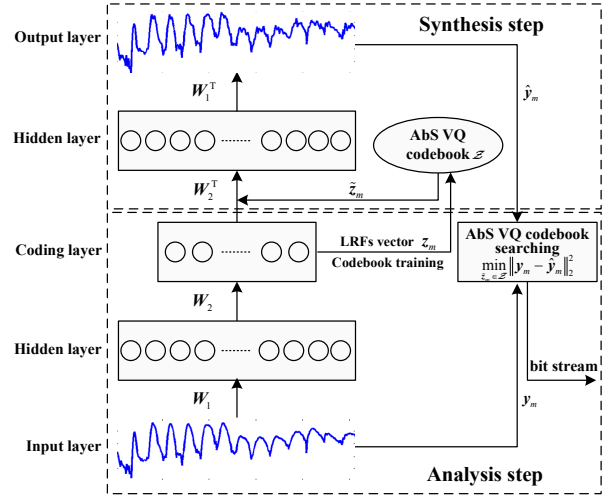}
% where an .eps filename suffix will be assumed under latex,
% and a .pdf suffix will be assumed for pdflatex; or what has been declared
% via \DeclareGraphicsExtensions.
\caption{Diagram of AbS VQ for LRFs vector.}
\label{fig_sim}
\end{figure}
\subsection{Suboptimal AbS VQ codebook searching}
The number of codewords in the AbS VQ codebook $\mathcal{Z}$ is usually very large, it is not practical to search the whole codebook because the computational complexity is too high. Consequently, we propose a low complexity suboptimal codebook searching technique, as is shown in Fig. 4. At first, we select some candidate codewords using the conventional VQ method to constitute a suboptimal codebook $\mathcal{Z}_s$, only the codewords in $\mathcal{Z}_s$ are chosen for synthesizing speech spectrum and calculating the log-spectral distortion. Hence, the computational complexity will be reduced dramatically since the number of codewords in $\mathcal{Z}_s$ is far less than that in $\mathcal{Z}$.

\begin{figure}[htp]
\centering
\includegraphics[width=3.1in]{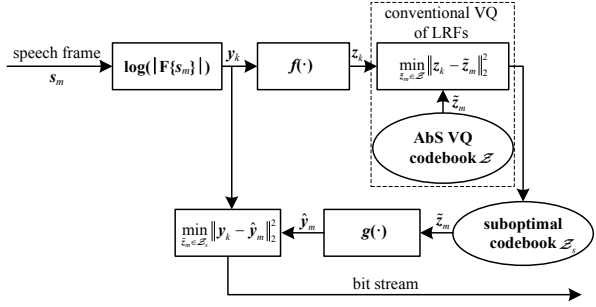}
% where an .eps filename suffix will be assumed under latex,
% and a .pdf suffix will be assumed for pdflatex; or what has been declared
% via \DeclareGraphicsExtensions.
\caption{Diagram of low complexity suboptimal codebook searching.}
\label{fig_sim}
\end{figure}

From the perspective of practical applications, we will use the split vector quantization technique (SVQ) as an alternative of direct VQ to further reduce the storage and computational complexity. In order to incorporate the SVQ technique into the AbS VQ framework, we will keep $J$ optimal candidate codewords while searching each sub-vector codebook, then we can make up the suboptimal codebook $\mathcal{Z}_s$ with different combination of these reserved sub-vector codewords. Therefore, $\mathcal{Z}_s$ will contain $J^D$ codewords if the original LRFs vector is divided into $D$ sub-vectors. Obviously, the AbS SVQ method will regress to be the conventional SVQ method if $J=1$.
\subsection{Bit allocation scheme for speech compression}
In order to remove the redundancy between adjacent speech frames, we will quantize the LRFs vector learned from $T$ consecutive speech frames using the AbS SVQ method, i.e. the joint log-magnitude spectral vector of $(N/2+1)$$\times$$T$ points is encoded as a whole, where $N/2+1$ is the dimension of the input vector for each speech frame. The frame length is 32 msec (256 samples) while the frame shift is 15 msec (120 samples)
, then we can get the bit allocation scheme as is shown in Tab. 1. In this bit allocation scheme, we can see that $\mathcal{Z}_s$ will contain $J^6$, $J^6$ codewords when the bit rate of speech coding is 2400 and 1200 bit/s, respectively.

\begin{table*}[htp]
\caption{Bit allocation of AbS SVQ scheme.}
\centering
{\begin{tabular}{ccccc}
\hlinewd{0.9pt}
Rate & {Bits/} & $T$& \multicolumn{1}{c}{Quantizaiton}\\
(bit/s) & Frame & (Frames)& \multicolumn{1}{c}{Scheme} \\\hline
2400 & 36 & 2 & \multicolumn{1}{c}{(12-12-12-12-12-12)-bit AbS SVQ} \\
1200 & 18 & 3 & \multicolumn{1}{c}{(9-9-9-9-9-9)-bit AbS SVQ}  \\
\hlinewd{0.9pt}
\end{tabular}}{}
\end{table*}

\subsection{Speech waveform reconstruction from spectrogram}
The step of spectrogram inversion in Deep Vocoder aims to estimate the discarded phase spectrum. Here we use the classic Griffin-Lim algorithm to complete this task for its simplicity. The Griffin-Lim algorithm iteratively estimates the phase spectrum via modified DFT and IDFT, and then couples it to the given magnitude spectrum resulting in a time-domain estimate of each speech frame \cite{MMSPSpeech:Min}. The speech waveform is finally reconstructed via an overlap-add procedure from the sequence of estimated speech frames.

\section{Experiments and Results}
\subsection{Dataset and evaluation metrics}
We conducted experiments on the widely used TIMIT corpus to evaluate the performance of Deep Vocoder. In the training stage, the complete TIMIT training set with 4620 utterances spoken by 462 speakers was used, the duration of training speech is $\thicksim$4 h. In the testing stage, we used the whole test set containing a total of 168 speakers and 1680 utterances, the duration of testing speech is $\thicksim$1.5 h. All the speech waveforms were downsampled to 8kHz. The speech signal was enframed to 256 samples using a hamming window, then the dimension of the input log-magnitude spectral vector for each speech frame is 129, i.e., $N/2+1=129$.

We will use three different metrics to evaluate the quality of compressed speech. The first is perceptual evaluation of speech quality (PESQ) \cite{ICASSP:Rix}, which is highly correlated with subjective evaluation scores and is always adopted as a standard objective measure. Another two metrics are frequency-weighted segmental SNR (fwsegSNRs) \cite{NJObjective:Quackenbush} and short-time objective intelligibility (STOI) \cite{TASLPAn:Taal}, which are also popular objective measures. PESQ and fwsegSNRs demonstrates the overall speech quality while the STOI measure illustrates the speech intelligibility. For both the metrics, higher score indicates better performance.
\subsection{DAE architecture and hyper-parameters setting}
An 11-layer deep autoencoder was trained on the TIMIT training set for analyzing and synthesizing speech signal, its architecture was set as 129$\times$$T$-2048-2048-1024-1024-72/54-1024-1024-2048-2048-$129$$\times$$T$ for considering both the performance of DNNs and the capacity of our hardware platform (Intel Xeon CPU(2.4GHz) and NVIDIA GeForce GTX Titan X GPU). The number of nodes in the DAE coding layer i.e. the dimension of LRFs vector is 72 and 54 when the bit rate of speech coding is 2400 bit/s and 1200 bit/s, respectively. We used sigmoid as the nonlinear activation function in our network for its bounded output. At the training stage, the size of each minibatch for RBM pre-training in each layer was 512, the learning rate was $1\times10^{-3}$ , the momentum rate was 0.99 and the number of iterations was 300. As for the fine-tuning stage, the learning rate was $1\times10^{-3}$ at first, then decreased by $1\times10^{-4}$ after each subsequent epoch, the momentum rate was 0.9, and the number of iterations was 1000.
\subsection{Evaluation of speech quality}
The experimental results at various bit-rates for the proposed Deep Vocoder and the conventional SQ- or SVQ-based vocoder are shown in Tables 2--4, in which the reuslts of conventional SQ method are marked in underline and the best results are highlighted in bold. It is worth noting that the AbS SVQ method and the conventional SVQ method are both equivalent when $J=1$. It is clearly illustrated that the proposed AbS VQ-based Deep Vocoder yields substantially higher fwsegSNRs, PESQ and STOI score than the conventional SQ and SVQ method, which demonstrates that the speech quality of Deep Vocoder is much better.

\begin{table*}[htb]
\begin{center}
\caption{Comparison on the fwsegSNRs with standard deviation (dB).}\label{tab:cap}
\begin{tabular}{cccccc}
\hlinewd{0.9pt}
Rate & Scalar &\multicolumn{3}{c}{AbS SVQ}& MELPe\\\cline{3-5}
{(bit/s)} & {Quantization} & J=1(SVQ) & J=2 & J=3 & Codec\\\hline
2400 & $\underline{10.06\pm0.23}$ & $12.46\pm0.34$ & $13.19\pm0.29$ & $\textbf{13.53}\pm\textbf{0.26}$ & $6.76\pm0.40$\\
1200 & $\underline{9.62\pm0.29}$ & $11.52\pm0.33$ & $12.09\pm0.29$ & $\textbf{12.10}\pm\textbf{0.39}$ & $5.51\pm0.41$\\
\hlinewd{0.9pt}
\end{tabular}
\end{center}
\end{table*}

\begin{table*}[htb]
\caption{Comparison on the PESQ score with standard deviation.}
\centering
{\begin{tabular}{cccccc}
\hlinewd{0.9pt}
Rate & Scalar &\multicolumn{3}{c}{AbS SVQ}& MELPe\\\cline{3-5}
{(bit/s)} & {Quantization} & J=1(SVQ) & J=2 & J=3 & Codec\\\hline
2400 & $\underline{2.92\pm0.09}$ & $3.06\pm0.09$ & $3.26\pm0.07$ & $\textbf{3.34}\pm\textbf{0.06}$ & $3.22\pm0.11$\\
1200 & $\underline{2.68\pm0.07}$ & $2.86\pm0.09$ & $3.04\pm0.08$ & $\textbf{3.08}\pm\textbf{0.11}$ & $3.00\pm0.12$\\
\hlinewd{0.9pt}
\end{tabular}}{}
\end{table*}

\begin{table*}[htb]
\caption{Comparison on the STOI score with standard deviation.}
\centering
{\begin{tabular}{cccccc}
\hlinewd{0.9pt}
Rate & Scalar &\multicolumn{3}{c}{AbS SVQ}& MELPe\\\cline{3-5}
{(bit/s)} & {Quantization} & J=1(SVQ) & J=2 & J=3 & Codec\\\hline
2400 & $\underline{0.77\pm0.01}$ & $0.86\pm0.01$ & $0.89\pm0.01$ & $\textbf{0.90}\pm\textbf{0.01}$ & $0.44\pm0.01$\\
1200 & $\underline{0.75\pm0.02}$ & $0.83\pm0.01$ & $0.86\pm0.01$ & $\textbf{0.87}\pm\textbf{0.01}$ & $0.28\pm0.01$\\
\hlinewd{0.9pt}
\end{tabular}}{}
\end{table*}

In detail, we can see that the advantage of AbS SVQ is significantly greater than SQ, this is because it fully exploits the correlation among the components of LRFs vector. Also, by comparing the performance of AbS SVQ to SVQ, we can find it very effective to use the analysis-by-synthesis mechanism in the quantization procedure of LRFs vector, which helped to improve the speech quality obviously. Moreover, $\mathcal{Z}_s$ approaches $\mathcal{Z}$ with the increasing of $J$, so the speech quality continues being improved. Specifically, the final improvement is impressive in the case of speech coding at 2400 bit/s, the average fwsegSNRs, PESQ and STOI score is approximately improved by 1.1dB, 0.3 and 4\%, respectively. Also, we can see that the outputting speech quality for Deep Vocoder with AbS VQ method is competitive with state-of-the-art MELPe codec. It should be noted that the fwsegSNRs and STOI measures are not very suitable for MELPe codec, because MELPe codec is not designed for minimizing the distortion in the spectral domain, it is designed to guarantee the overall speech quality. However, PESQ is usually adopted for evaluating the performance of MELPe codec.

As mentioned before, larger $J$ will generate more codewords in $\mathcal{Z}_s$, hence, to make the trade-off between speech quality and computational complexity, we will specify $J=3$ in the case of speech coding at 2400, 1200 bit/s, respectively.
Fig.5 shows the spectrograms of the original speech and the reconstructed speech via Deep Vocoder for a typical TIMIT utterance. We can see that the frequency formant structure and harmonic structure are both well preserved for the reconstructed spectrogram via Deep Vocoder, which illustrates that the coded speech sounds close to the original speech. Inevitably, the magnitude spectrum of few unvoiced speech segments is smeared due to the quantization procedure of LRFs vector, which causes synthesis artifacts and slightly degrades the articulation of reconstructed speech. However, the reconstructed speech for Deep Vocoder is free of the harsh synthetic sounds arisen in many model-based vocoders, such as CELP, MELP codec.
\begin{figure}[htp]
\centering
\includegraphics[width=3.1in]{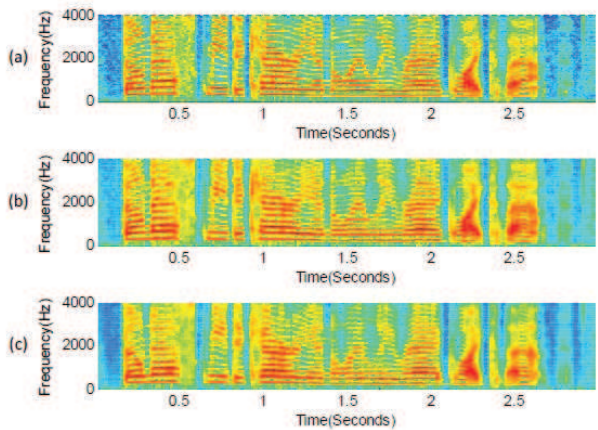}
% where an .eps filename suffix will be assumed under latex,
% and a .pdf suffix will be assumed for pdflatex; or what has been declared
% via \DeclareGraphicsExtensions.
\caption{Spectrograms comparison of the TIMIT utterance ``Don't ask me to carry an oily rag like that''. (a) original speech; (b) reconstructed speech
via Deep Vocoder at 2400 bit/s; (c) reconstructed speech via Deep Vocoder at 1200 bit/s.}
\label{fig_sim}
\end{figure}
\section{Conclusion}
We introduce Deep Vocoder, a DAE-based framework for speech compression at 2400 bit/s and 1200 bit/s in this paper. We propose an analysis-by-synthesis vector quantization approach for low bit rate Deep Vocoder. The objective of AbS VQ is changed to minimize the perceptual spectral reconstruction distortion rather than the distortion of LRFs vector. A suboptimal codebook searching technique is also proposed for practical implication. Experimental results show that the speech quality is substantially improved when compared to the output of conventional SQ- or VQ-based codec.

It is expected that the performance of Deep Vocoder can be improved further. Instead of using the DAE model to analyze and synthesize speech signal, this procedure could be optimized by utilizing some new generative models, such as variational autoencoder (VAE) or generative adversarial networks (GANs)\cite{arXiv:Kingma}, \cite{arXiv:Goodfellow}. Also, the Griffin-Lim-based speech waveform synthesizing method is also ripe for improvement since the Griffin-Lim outputs may suffer from audible artifacts. We are planning to explore high-quality spectrogram inversion method using deep neural networks in the future.

\bibliographystyle{IEEEbib}

\bibliography{icme2019template}

\end{document}